\documentclass[a4paper]{jpconf}
\usepackage{graphicx}
\begin{document}
\title{3D hydrodynamical model atmospheres: a tool to correct radial velocities and parallaxes for Gaia.}

\author{A. Chiavassa$^1$, L. Bigot$^2$, F. Th\'evenin$^2$, R. Collet$^3$, G. Jasniewicz$^4$, Z. Magic$^3$, M. Asplund$^3$}

\address{$^1$ Institut d'Astronomie et d'Astrophysique, Universit\'e Libre de Bruxelles, CP. 226, Boulevard du Triomphe, B-1050 Bruxelles, Belgium}
\address{$^2$ Universit\'e de Nice Sophia-antipolis, CNRS Laboratoire Cassiop\'ee, Observatoire de la C\^{o}te dÕAzur,
B.P. 4229, 06304 Nice Cedex 4, France}
\address{$^3$ Max-Planck-Institut f\"{u}r Astrophysik, Karl-Schwarzschild-Str. 1, Postfach 1317, D-85741 Garching b. M\"{u}nchen, Germany}
\address{$^4$ LUPM, Laboratoire Univers et Particules, Universit\'{e} de Montpellier II, CNRS, Place Eug\'{e}ne Bataillon, 34095 Montpellier Cedex 05, France}

\ead{achiavas@ulb.ac.be}

\begin{abstract}
Convection plays an essential role in the emerging intensity for many stars that will be observed by Gaia. Convective-related surface structures affect the shape, shift, and asymmetry of absorption lines, the phocentric and photometric variability causing bias in Gaia measurements. Regarding the importance of Gaia mission and its goals, it is mandatory to have the best models of the observed stars. 3D time-dependent hydrodynamical simulations of surface convection are crucial to model the photosphere of late type stars in a very realistic way. These simulations are an important tool to correct the radial velocities and better estimates the parallaxes and photometric variability.
\end{abstract}

\section{Introduction}

The main goal of the Gaia mission \cite{2001A&A...369..339P,2008IAUS..248..217L} is to determine high-precision astrometric parameters (i.e., positions,
parallaxes, and proper motions) for one billion objects with apparent magnitudes in the range $5.6 \le V \le 20$ and kinematic velocities of about 100 millions of stars with a precision of $\sim$1 km.s$^{-1}$ up to  $V \le 13$. These data along with multi-band and multi-epoch photometric data will allow to reconstruct the formation history, structure, and evolution of the Galaxy.\\
Convection plays a crucial role in the formation of spectral lines and deeply influences the shape, shift, and asymmetries of lines in late type stars which will represent most of the objects that will be observed by Gaia. In addition to this, granulation-related variability that is considered as "noise" must be quantified in order to better characterize any resulting systematic error on the parallax and photometric determinations. \\

Realistic modelling of stellar atmospheres is therefore crucial for a better interpretation of future Gaia data and, in this context, three-dimensional radiative-hydrodynamical models are needed for a quantitative correction of the radial velocities (few hundreds km.s$^{-1}$) for all the stars observed and, in evolved stars, for the determination of the photocenter positions.

\section{Surface convection simulations and radiative transfer calculations}

We adopted here time-dependent, three-dimensional (3D), radiative-hydrodynamical (RHD) surface convection simulations of different stars across the Hertzsprung-Russell diagram (Tab.~\ref{simus}). The simulations have been carried out using:

\begin{itemize}
\item The \textsc{box-in-a-star} setup with {\sc Stagger-Code}\footnote{www.astro.ku.dk/{\textasciitilde}aake/papers/95.ps.gz}. These models cover
only a small section of the surface layers atop the deep convection
zone, and the numerical box includes a number of convective cells
proportional to the surface gravity. They have constant gravity, the lateral boundaries are periodic, and the radiation transport module relies on a Feautrier scheme applied to long characteristics.
\item The \textsc{star-in-a-box} setup with CO$^5$BOLD code \cite{2002AN....323..213F,Freytag2008A&A...483..571F,chiavassa2011}. These simulations cover the
whole convective envelope of the star and have been used to model red supergiant (RSG) stars \cite{2002AN....323..213F,chiavassa2011}
and Asymptotic giant branch stars \cite{Freytag2008A&A...483..571F} so far. The computational domain is an equidistant cubic grid in all directions, and the same
open boundary condition is employed for all side of the computational box.
\end{itemize}

The transition between the box-in-a-star and star-in-a-box occurs around $\log g \sim 1$, when the influence of sphericity and the ratio between granule size versus
stellar diameter become important; the star-in-a-box global models are then needed, but those are highly computer-time demanding and difficult to run so that there are only very few models available so far. \\

We used the 3D pure-LTE radiative transfer code {\sc Optim3D} \cite{2009A&A...506.1351C} to compute spectra and intensity maps from
the snapshots of the RHD simulations listed in Table~\ref{simus}. The code takes into account the
Doppler shifts due to convective motions. The radiative
transfer equation is solved monochromatically using extinction coefficients
pre-tabulated with the same chemical compositions as the RHD simulations and 
using the same extensive atomic and molecular opacity data as the latest generation of
MARCS models \cite{2008A&A...486..951G}. We assumed a zero
micro-turbulence since the velocity fields inherent in 3D models
are expected to self-consistently and adequately account for
non-thermal Doppler broadening of spectral lines.

\begin{table}[h]
\caption{\label{simus}3D hydrodynamical model atmospheres used in this work. The symbols refer to Fig.~\ref{chilocal}, \ref{chiglobal}, \ref{amplitude}, and \ref{width}.}
\begin{center}
\begin{tabular}{c c c c c c c c c c}
\br
Stellar & $T_{\rm{eff}}$ & $\log g$ & [Fe/H] & Mass & $\rm{R}_{\star}$&  $\Delta \lambda_{\rm{FeI}}$ &  $\Delta \lambda_{\rm{CaII}}$ & Ref. & Symbol \\
type &  $[$K$]$     &    &  &    $[M_\odot]$&   $[\rm{R}_\odot]$  & [km.s$^{-1}$] & [km.s$^{-1}$] & & \\
\mr
  \multicolumn{9}{c}{Local simulations with {\sc Stagger-Code}}  \\
   \multicolumn{9}{c}{ }\\
K giant & 4700 & 2.2 & $\textrm{ }$$\textrm{ }$0.0 & $\dots$ & $\dots$ & $-$0.36  & $+$0.29 &   \cite{2007A&A...469..687C} & $\textrm{ }$\color{black}\fullcircle \\
K giant & 4720 & 2.2 & $-$1.0 & $\dots$ & $\dots$ & $-$0.45   & $+$0.23 & \cite{2007A&A...469..687C} & \color{red}\fulltriangle \\
K giant & 5035 & 2.2 & $-$2.0 & $\dots$ & $\dots$ & $-$0.58   & $+$0.25 & \cite{2007A&A...469..687C} & \color{yellow}\fulltriangledown \\
K giant &5130 & 2.2 & $-$3.0 & $\dots$ & $\dots$ & $-$0.28   & $+$0.31 & \cite{2007A&A...469..687C} & \color{green}\fulldiamond \\
K giant &4630 & 1.6 & $-$3.0 & $\dots$ & $\dots$ & $-$0.22   & $+$1.55 & \cite{2009MmSAI..80..719C} & \color{lightblue}\fullsquare \\ 
F star    & 6500 & 4.0 & $\textrm{ }$$\textrm{ }$0.0 & $\dots$ & $\dots$ & $-$0.75 & $+$3.4 & \cite{magic2011} & \color{blue}\fullstar \\
\mr
  \multicolumn{9}{c}{Global simulations with CO$^5$BOLD} \\
  \multicolumn{9}{c}{ }\\
RSG & 3430 & $-$0.35 & $\textrm{ }$$\textrm{ }$0.0 & 12 & 846 &  $+$0.75 & $-$1.89 & \cite{chiavassa2011}  & $\textrm{ }$\fullcircle \\
RSG &3660 & 0.02 & $\textrm{ }$$\textrm{ }$0.0 & 6 & 386 & $+$2.80  & $-$7.95 & \cite{chiavassa2011} & \color{red}\fullstar \\
\br
\end{tabular}
\end{center}
\end{table}

\section{Kinematic radial velocities and correction of convective shifts}

A convenient and usual way to estimate kinematic radial velocity of a star is commonly made from a measurement of its spectroscopic
radial velocity. Gaia is provided with a dedicated radial velocity spectrometer (RVS, \cite{2004MNRAS.354.1223K}) with a resolving power at best of 11\,500 centered on Ca $\mathrm{II}$ triplet in the spectral band from 8480 to 8750 \AA . It is well known that photospheric absorption lines in many stars are blueshifted as a result of convective motions in the stellar atmosphere. In fact, hot bright and rising (i.e., blueshifted) convective elements contribute with more photons than cool dark shrinking gas: as a consequence, the absorption lines appear blueshifted \cite{1982ARA&A..20...61D, 2002ApJ...566L..93A}. This effect is one of the systematic errors of spectroscopic measurements with RVS \cite{2003A&A...401.1185L, 2009MmSAI..80..622A} and will be taken into account by the analysis pipeline of Gaia using 3D RHD simulations \cite{2011EAS....45..195J}, which naturally account for turbulent motions and therefore do not need the use of the traditional micro and macro turbulence used in hydrostatic models.\\

\begin{figure}[h]
\begin{minipage}{18pc}
\includegraphics[width=19pc]{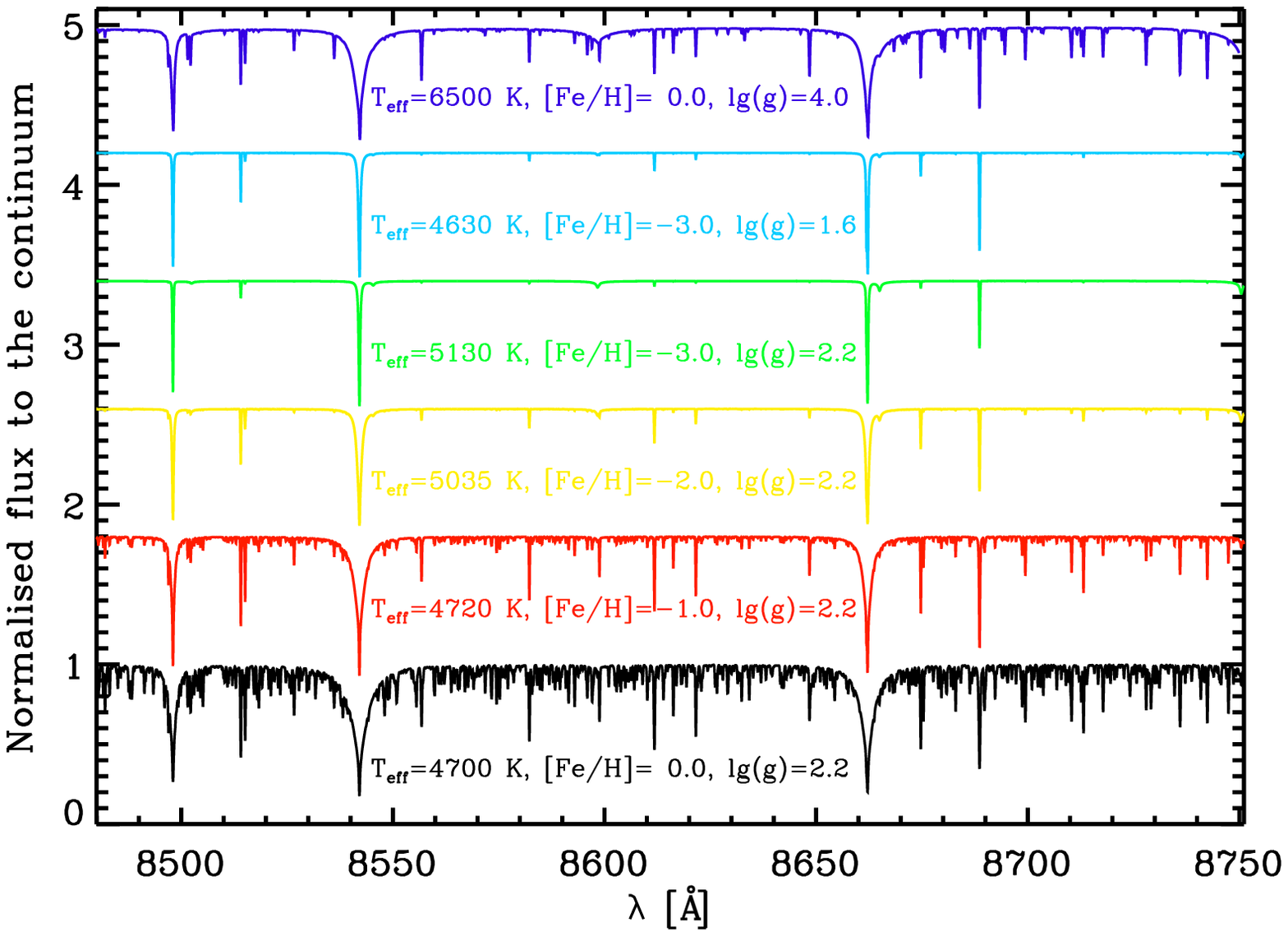}
\caption{\label{spectralocal}Spectra in the Gaia-RVS domain for the local simulations of Tab.~\ref{simus}. An offset has been added to the colored curves.}
\end{minipage}\hspace{1pc}%
\begin{minipage}{18pc}
\includegraphics[width=19.5pc]{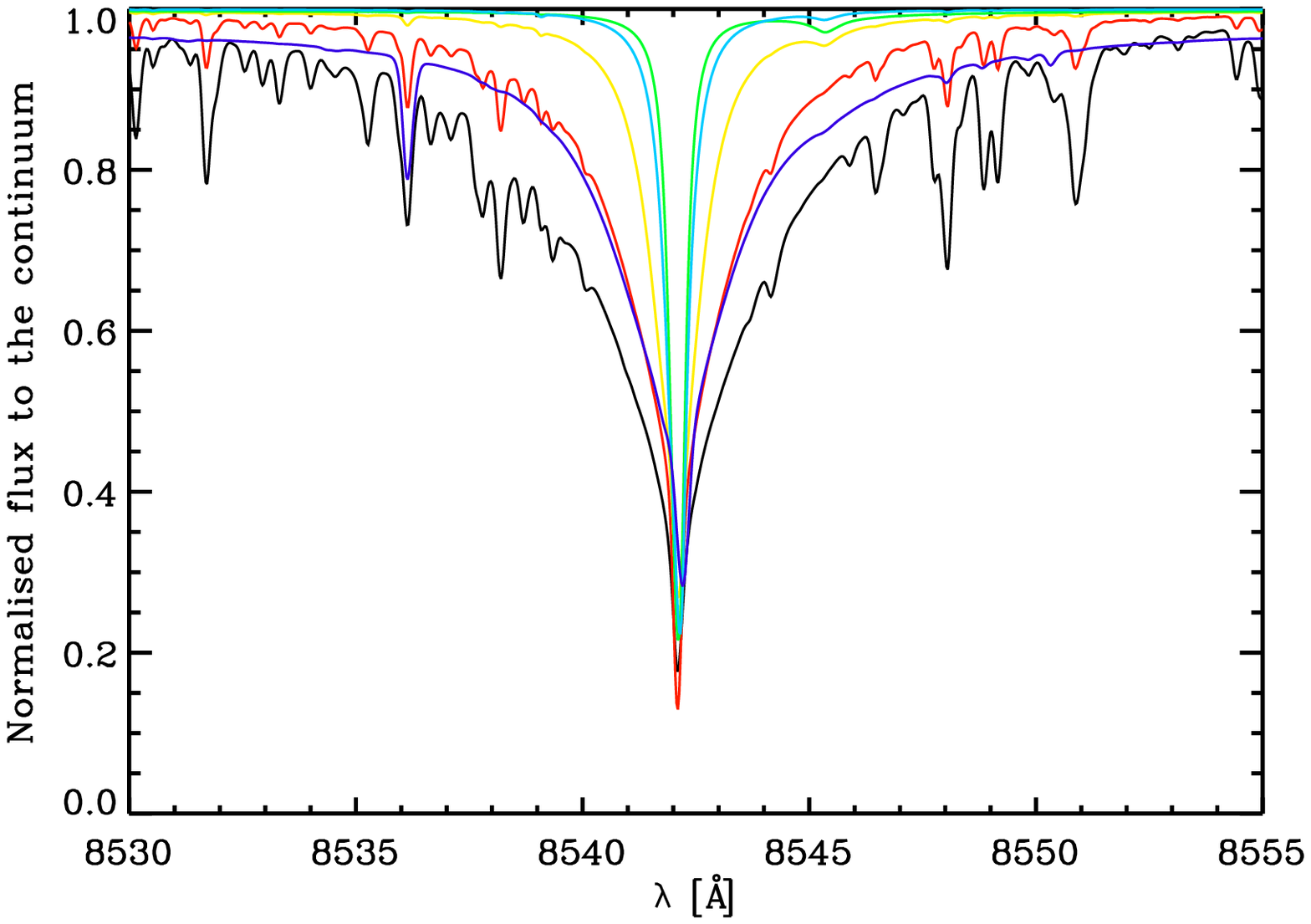}
\caption{\label{spectralocalzoom}Enlargements of Fig.~\ref{spectralocal} around a Ca $\mathrm{II}$ line.}
\end{minipage} 
\end{figure}

\begin{figure}[h]
\begin{minipage}{18pc}
\includegraphics[width=19pc]{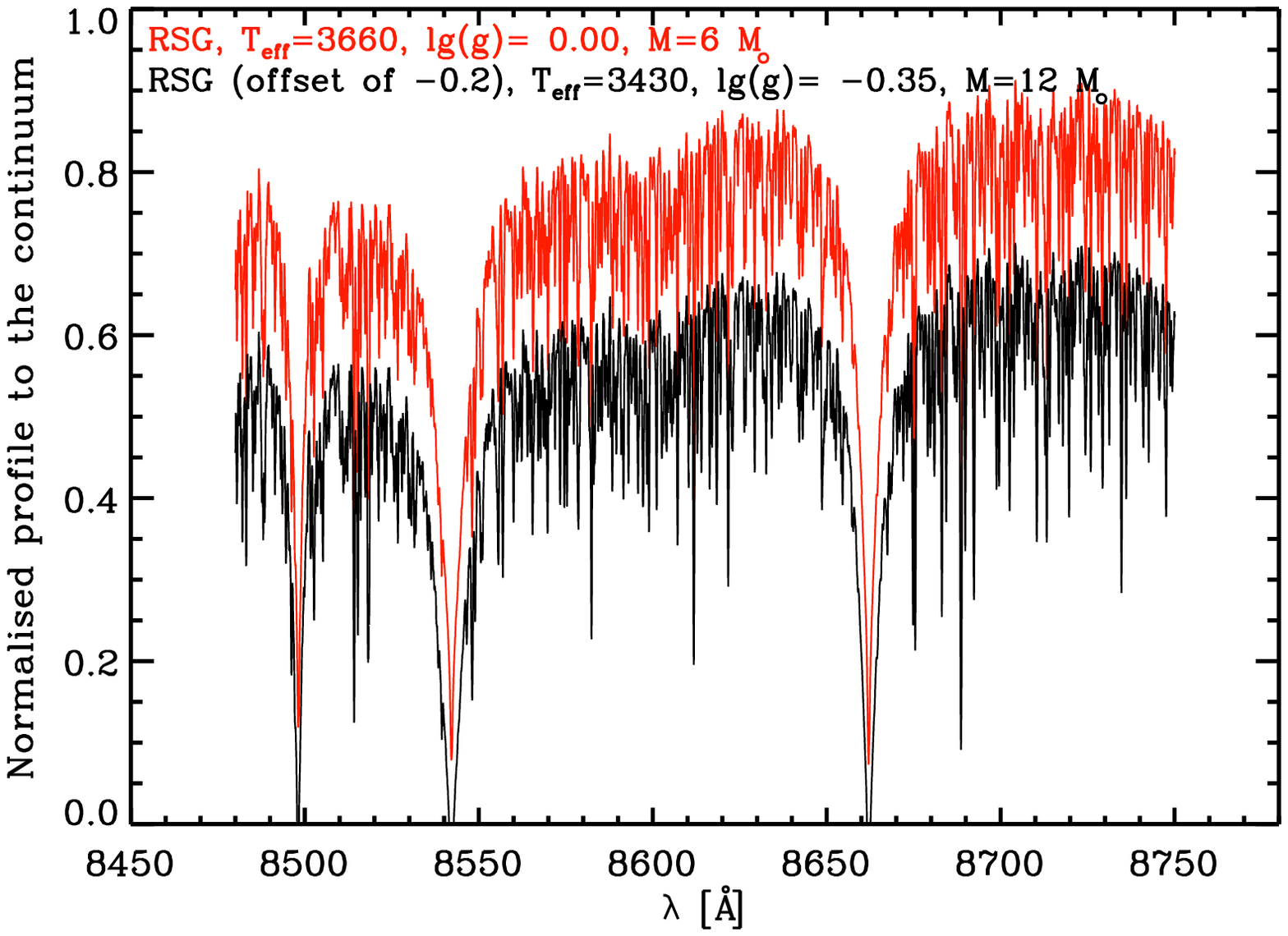}
\caption{\label{spectraglobal}Spectra in the Gaia-RVS domain for the global simulations of Tab.~\ref{simus}.}
\end{minipage}\hspace{2pc}%
\begin{minipage}{18pc}
\includegraphics[width=19pc]{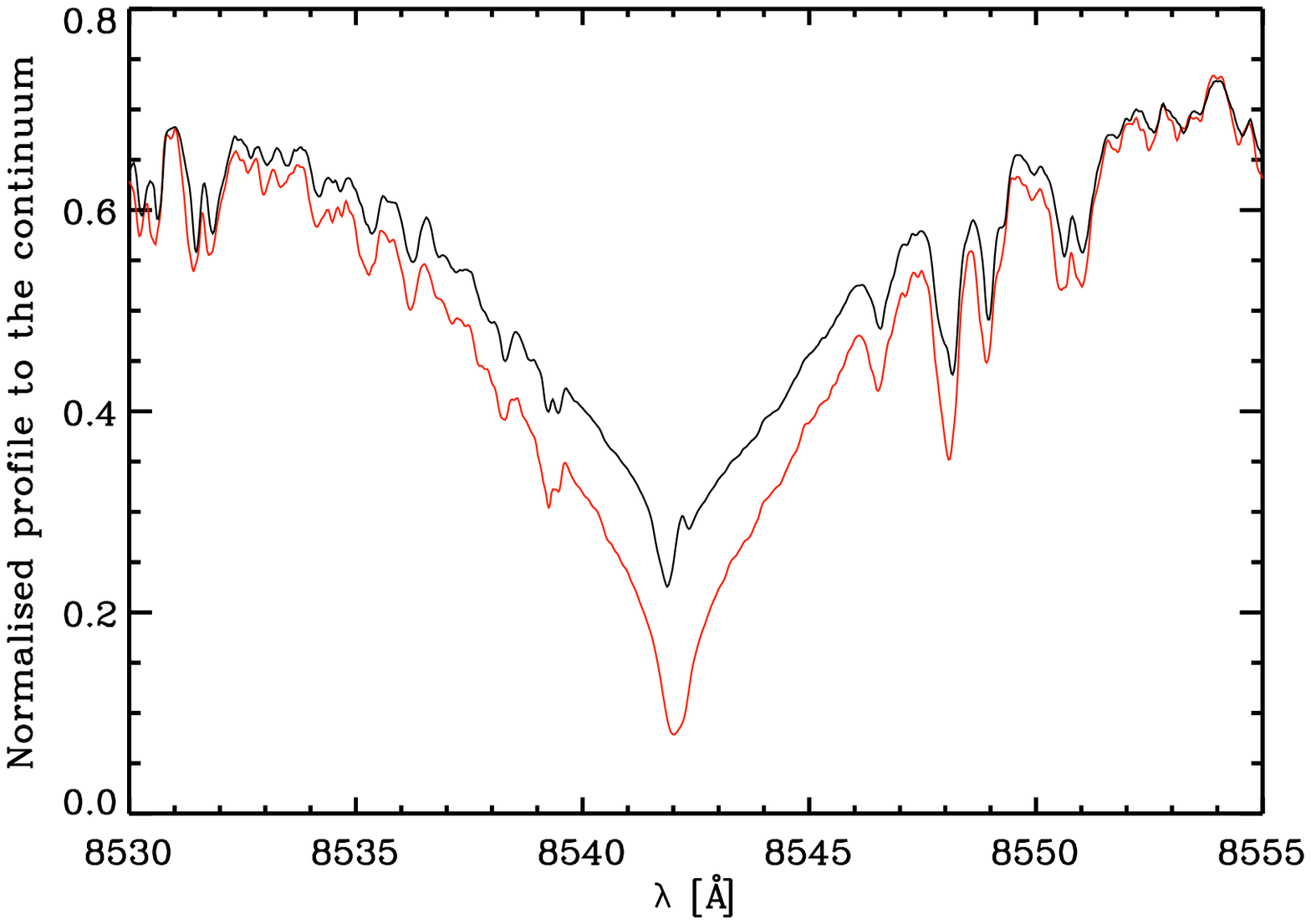}
\caption{\label{spectraglobalzoom}Enlargements of Fig.~\ref{spectraglobal} around a Ca $\mathrm{II}$ line.}
\end{minipage} 
\end{figure}

We computed synthetic spectra in RVS domain using {\sc Optim3D} for all the models reported in Tab.~\ref{simus}. The spectra were computed along rays of four $\mu-$angles $[0.88, 0.65, 0.55, 0.34]$ and four $\phi-$angles $[0^\circ, 90^\circ, 180^\circ, 270^\circ]$, after which we performed a disk integration and a temporal average over all selected snapshots. It is important that the total time covered by the simulations is such that there are no trends in the lineshift if a subset of snapshots is used for the calculations. This has been verified by \cite{2010ApJ...725L.223R} for the simulation with $T_{\rm{eff}}$=4630 and $\log g$=1.6 in Tab.~\ref{simus}. Ramirez et al. \cite{2010ApJ...725L.223R} found that the sequence was long enough to ensure that the results were statistically significant and the comparison with observations is also very good. However, the other K giant models we haven't done any comparisons with observations yet.\\ 
Figures~\ref{spectralocal} and \ref{spectralocalzoom} show the results for the local simulations and Fig.~\ref{spectraglobal} and \ref{spectraglobalzoom} the results for the global ones. It is noticeable how the Ca $\mathrm{II}$ triplet ($\lambda=$8498.02, 8542.09, 8662.14 \AA) is well visible in all the stars, even at low metallicity.\\

\begin{figure}[h]
\begin{minipage}{19pc}
\includegraphics[width=19pc]{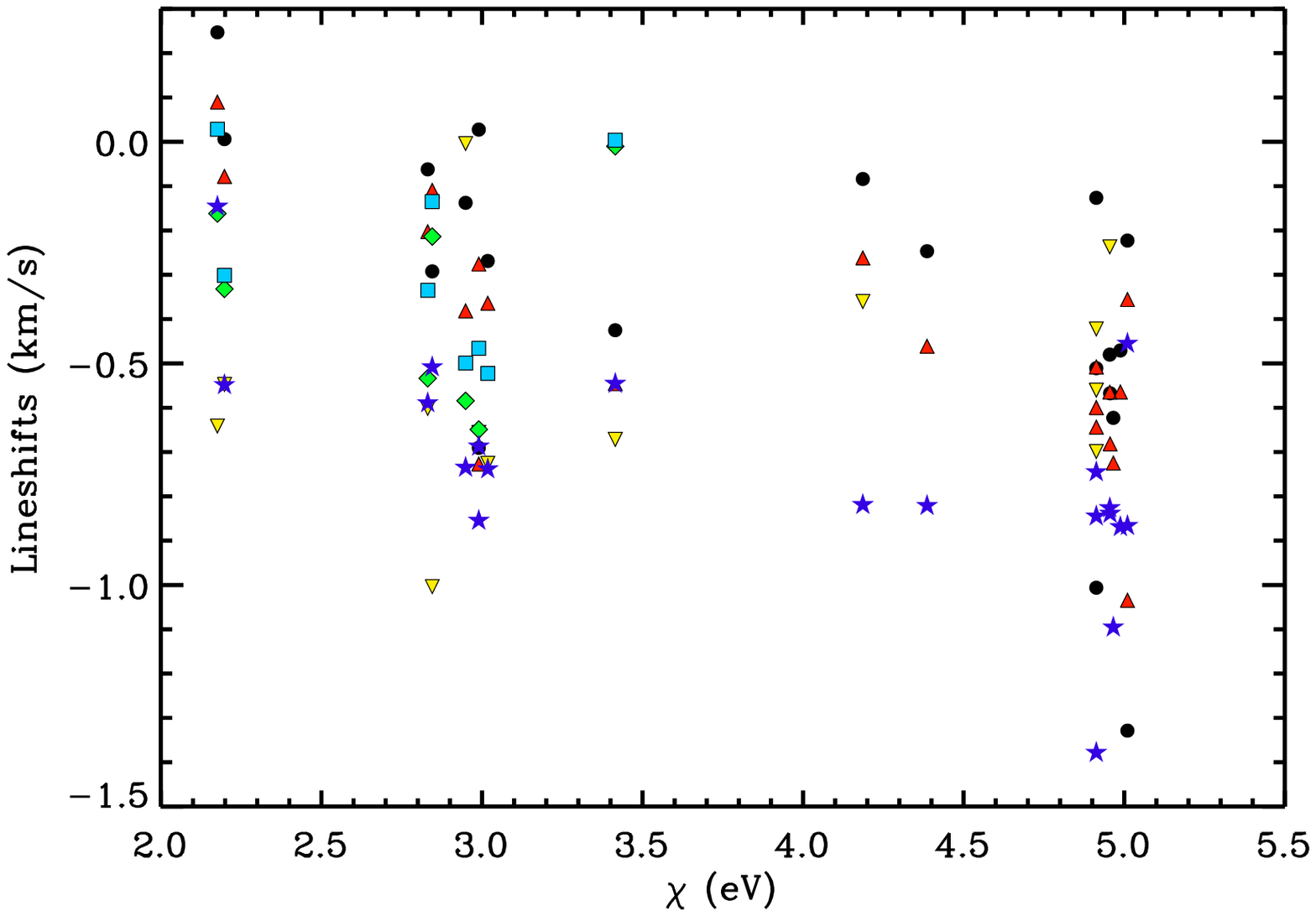}
\caption{\label{chilocal}Convective shifts of 20 Fe $\mathrm{I}$ from \cite{2006MNRAS.372..609B} as a function of excitation potentials for local simulations of Tab.~\ref{simus}. For metalpoor stars, only visible lines in the spectra are reported.}
\end{minipage}\hspace{0.5pc}%
\begin{minipage}{19pc}
\includegraphics[width=19pc]{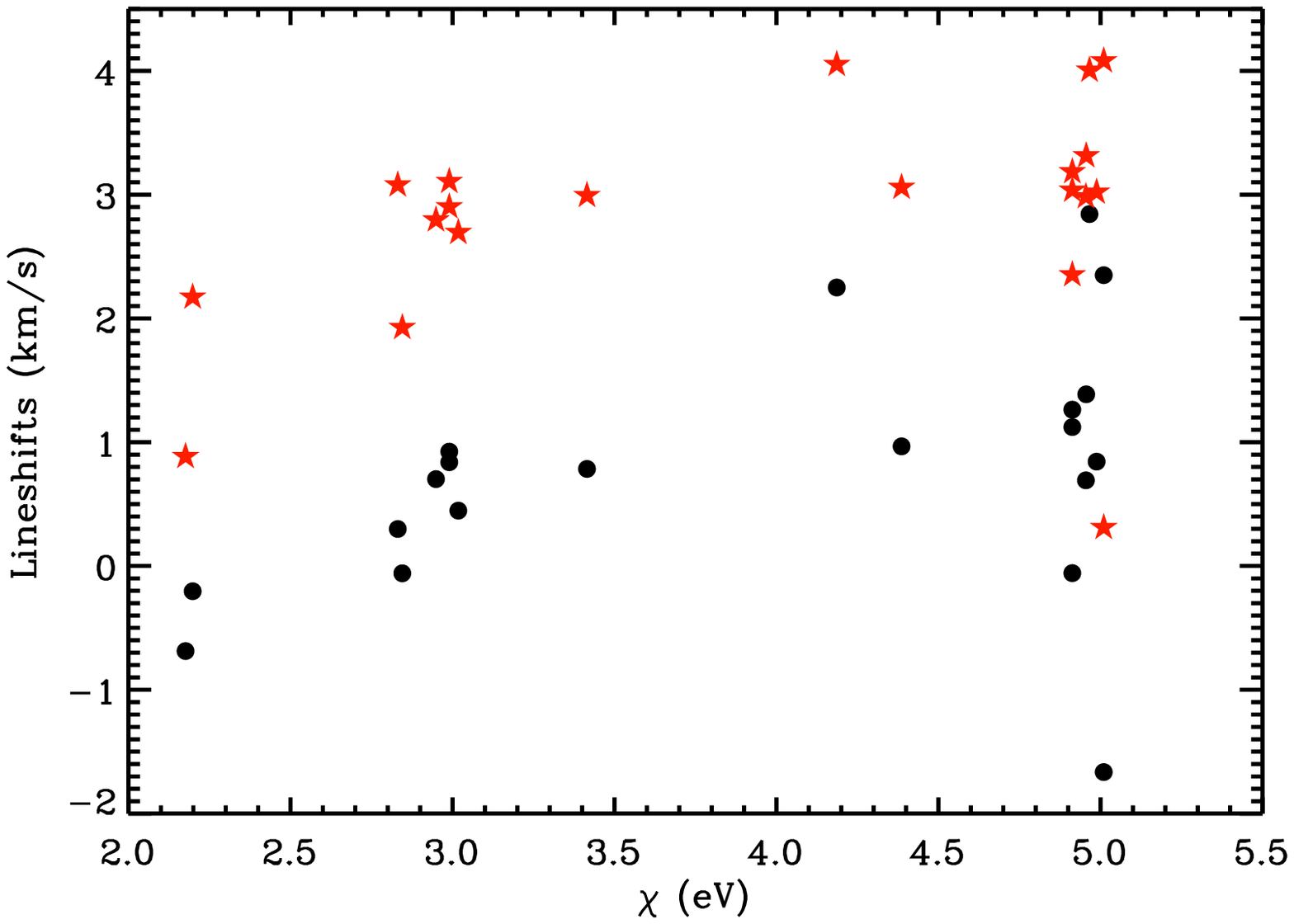}
\caption{\label{chiglobal}Convective shifts of 20 Fe $\mathrm{I}$ from \cite{2006MNRAS.372..609B} as a function of excitation potentials for global simulations of Tab.~\ref{simus}.}
\end{minipage} 
\end{figure}

Following the work of \cite{2008sf2a.conf....3B}, we determined the position, the depth and the width of Ca $\mathrm{II}$ triplet and of 20 Fe $\mathrm{I}$ lines selected to be non blended lines and to cover the spectral domain of the RVS \cite{2006MNRAS.372..609B}. For this preliminary work, we fitted the absorption lines with a gaussian function. \\
Figure~\ref{chilocal} shows that Fe $\mathrm{I}$ are mostly blueshifted (except for few cases with low excitation potential). The convective lineshift (Tab.\ref{simus}, $\Delta \lambda_{\rm{FeI}}$) appears stronger for high excitation lines and it ranges, in average, from -0.22 km.s$^{-1}$ (K giant with [Fe/H]=-3.0) to -0.75 km.s$^{-1}$ (F star). The amplitude of the shift increases when
going from K giants to the F star, as a consequence of the more vigorous convective motions. Moreover, the Ca $\mathrm{II}$ triplet lines are redshifted because these lines are formed in the upper part of the photosphere where the granulation pattern is reversed. The lineshift is particularly strong in the case of the F star (Tab.\ref{simus}, $\Delta \lambda_{\rm{CaII}}$).

\begin{figure}[h]
\begin{minipage}{18pc}
\includegraphics[width=19pc]{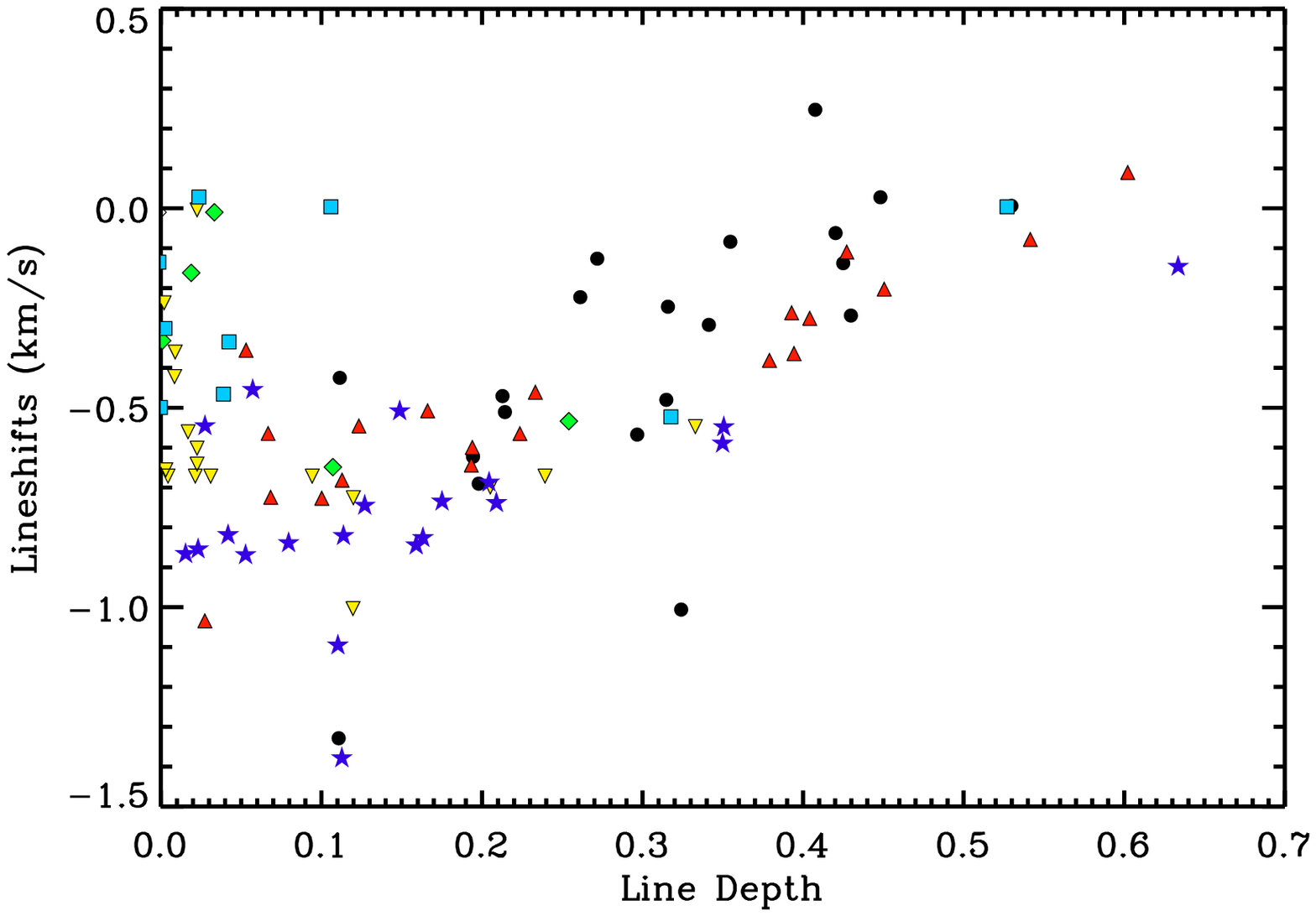}
\caption{\label{amplitude}Same as in Fig.~\ref{chilocal} but as a function of line depth.}
\end{minipage}\hspace{0.8pc}%
\begin{minipage}{18pc}
\includegraphics[width=19pc]{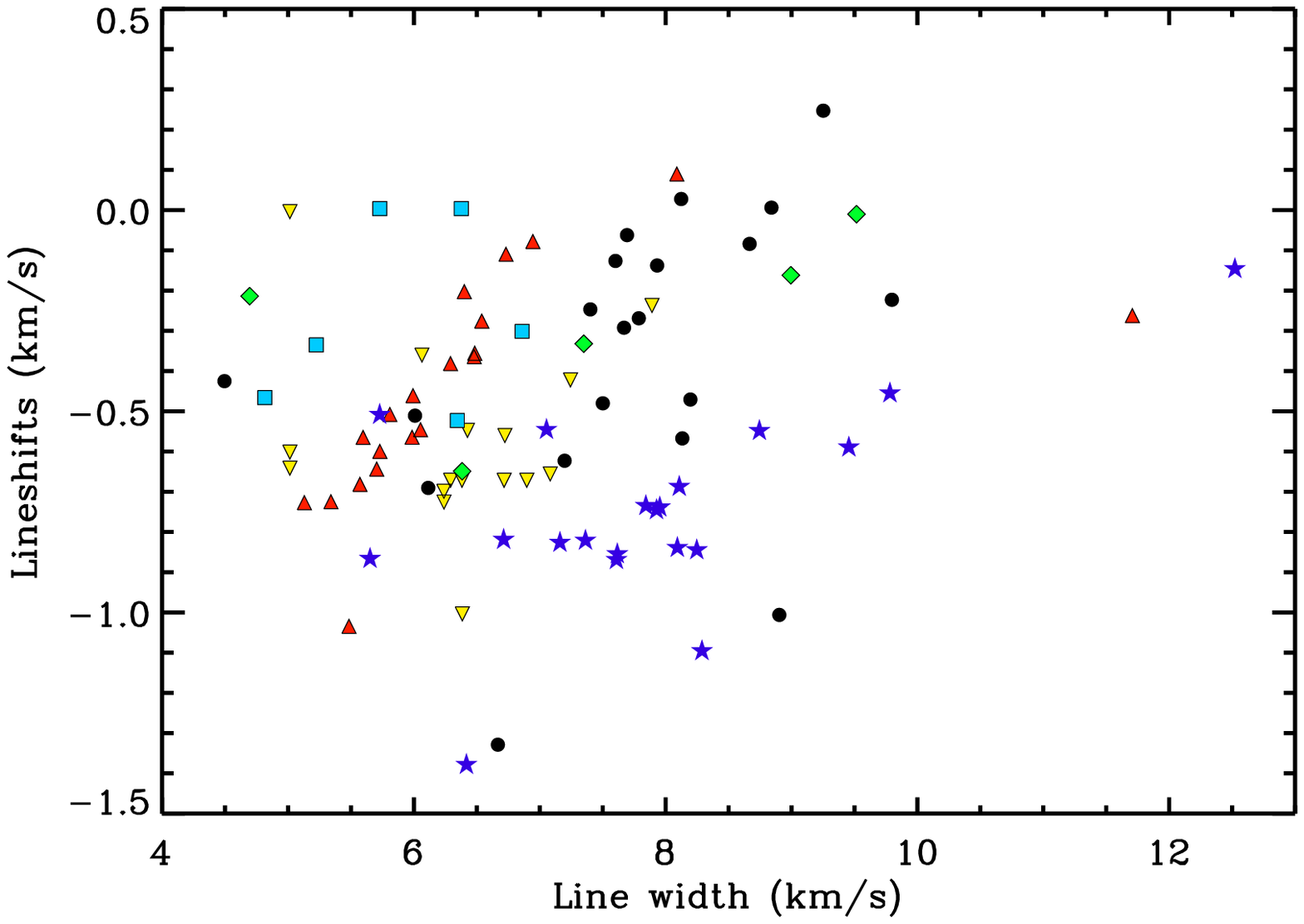}
\caption{\label{width}Same as in Fig.~\ref{chilocal} but as a function of line width.}
\end{minipage} 
\end{figure}

The lineshifts are not the same for all the lines in the RVS spectra. The precise amount of shift depends on the strength of the absorption line (and hence on the stellar metallicity), since different lines are formed at different atmospheric depths and thus experience different granulation contrasts and convective velocities. In fact, Fig.~\ref{amplitude} displays that there is a correlation between lineshifts and line depth indicating that weaker line (high excitation potential) are in general more shifted than deep lines (low excitation potential). This picture is also supported by Fig.~\ref{chilocal}. In addition to this, Fig.~\ref{width} shows that there is not apparent connection with the line width as found by \cite{2000A&A...359..729A} for the Sun.\\

RSGs show a completely different behavior (Fig.~\ref{chiglobal}) with strong redshifted Fe $\mathrm{I}$ up to 2.80 km.s$^{-1}$ and blueshifted Ca $\mathrm{II}$ up to 8 km.s$^{-1}$ (Tab.\ref{simus}). The reason is likely connected with the reversed-C shape of line bisectors observed by \cite{2008AJ....135.1450G} on the prototypical RSG Betelgeuse and caused by a peculiar and vigorous convection but further investigations are needed and still in progress with RHD simulations. \\

In conclusion, velocity shifts of F and K giant stars are of the order of RVS accuracy ($\sim$1 km.s$^{-1}$), and they are even larger for RSG stars.

\section{Photometric and photocentric variabilities and impact on parallaxes}

Massive evolved stars like RSG stars give rise to  large granules comparable to the stellar radius in the $H$ and $K$ bands, and an irregular pattern in the optical region \cite{2010A&A...515A..12C}. These  surface inhomogeneities vary with time and may affect strongly the photmetric and astrometric measurements of Gaia.\\

\begin{figure}[h]
  \begin{tabular}{c}
   \includegraphics[width=0.49\hsize]{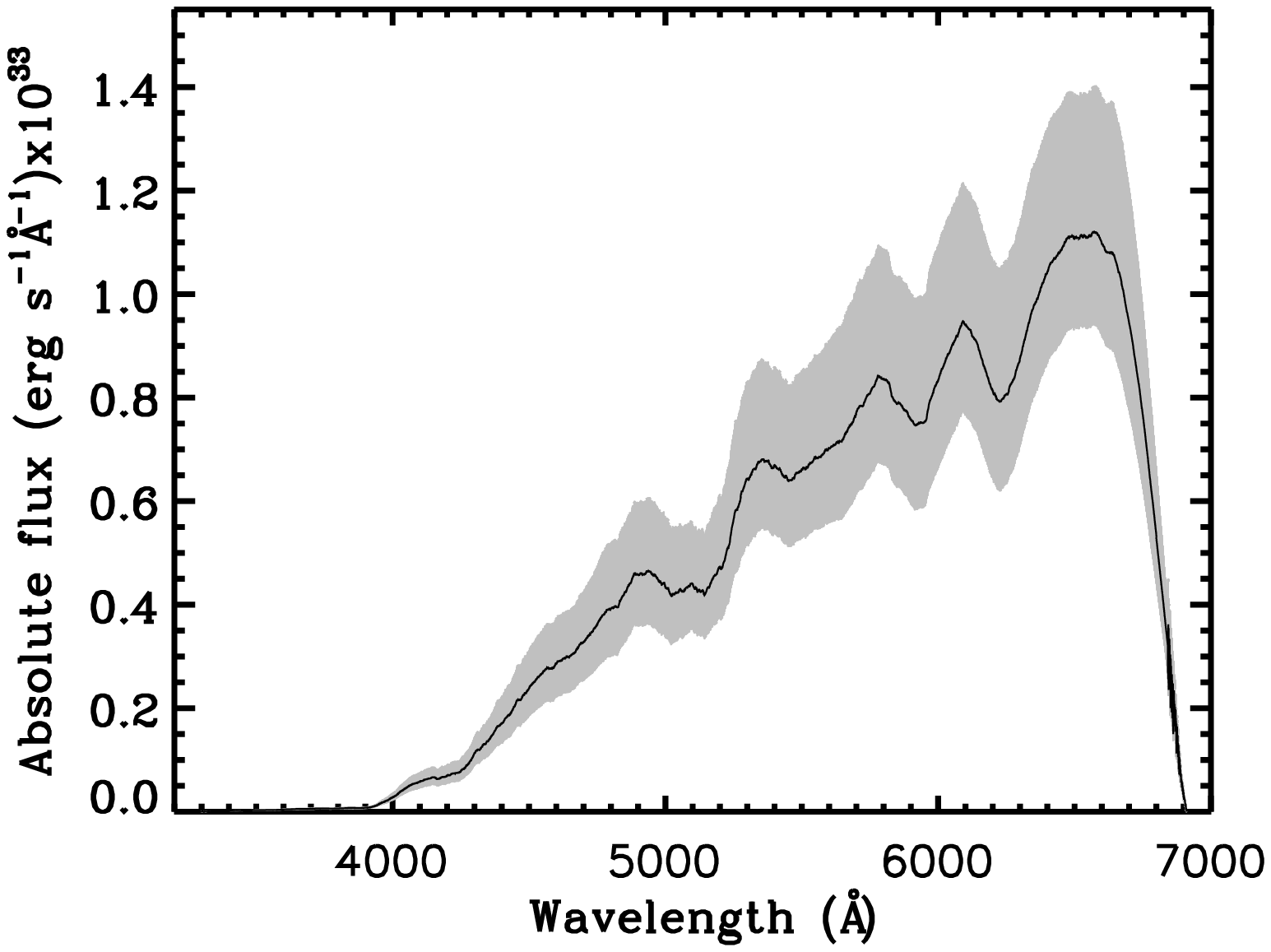}
   \includegraphics[width=0.48\hsize]{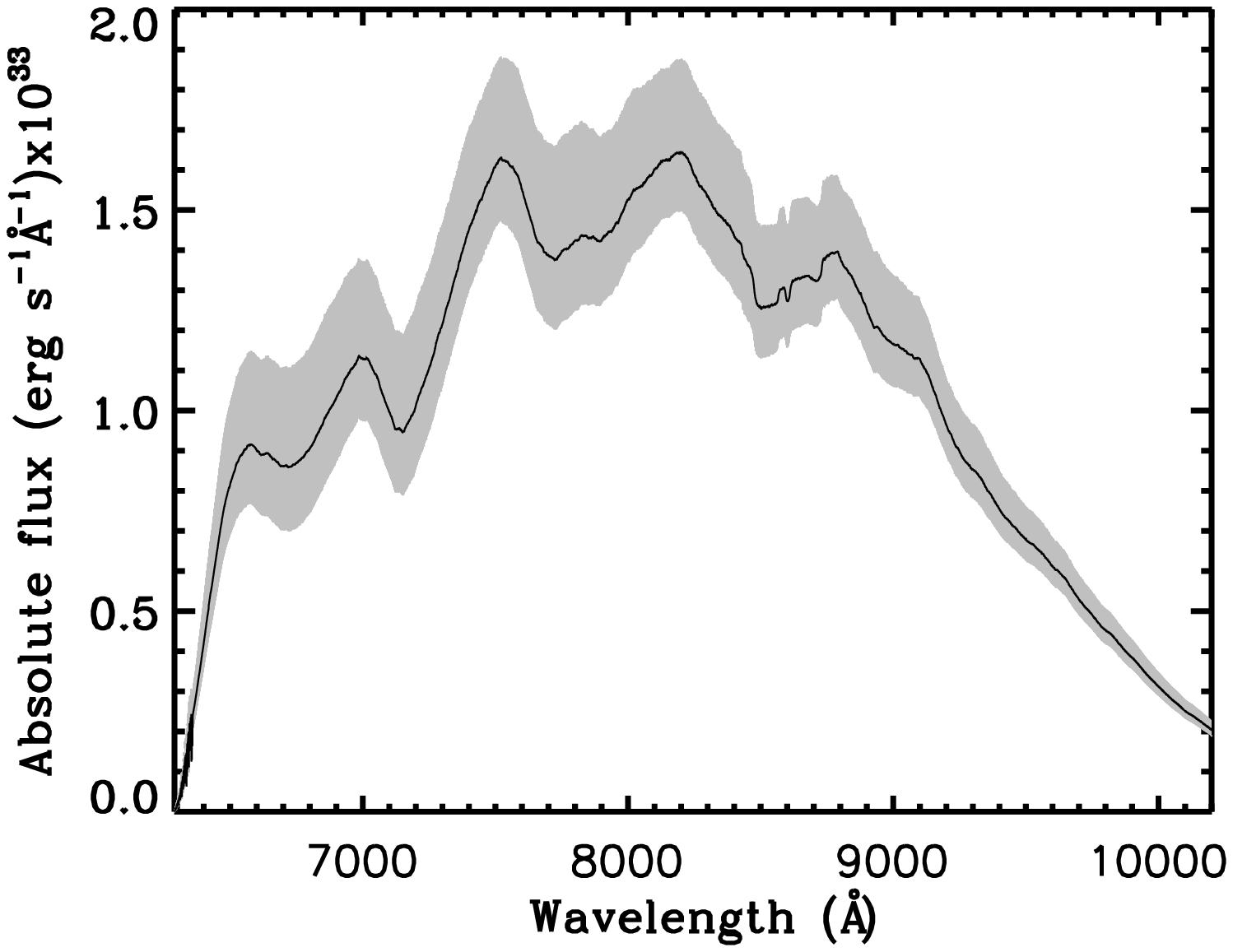}
\end{tabular}
     \caption{\label{blue}Spectral fluctuations in the blue (\emph{left}) and red (\emph{right}) Gaia
        photometric band \cite{2010A&A...523A..48J} for a RSG simulation. The black curve is the average flux over $\sim$5 years covered by the
simulation, while the "grey shade" denotes the maximum and minimum fluctuations. The spectra have been smoothed to the Gaia photometric resolution of $R\sim50$ \cite{2008PhST..133a4010T}. Figures taken from \cite{2011A&A...528A.120C}.}
\end{figure}

Figure~\ref{blue} displays the variability of a RSG during a period comparable with the Gaia mission in the blue and red photometric bands of Gaia. The photometric
system of Gaia will be used to characterize the star's effective temperature, surface gravity and metallicity \cite{2008PhST..133a4010T}. The simulations show fluctuations up to 0.28 mag in the blue and up to 0.15 mag in the red bands over 5 years. The photometric fluctuations of RGSs are not negligible and the uncertainties on [Fe/H], $T_{\rm{eff}}$ and $\log g$ for these stars with $G<15$ should be revised upwards for RSGs due to their convective motions \cite{2011A&A...528A.120C}.\\

\begin{figure}[h]
\includegraphics[width=20pc]{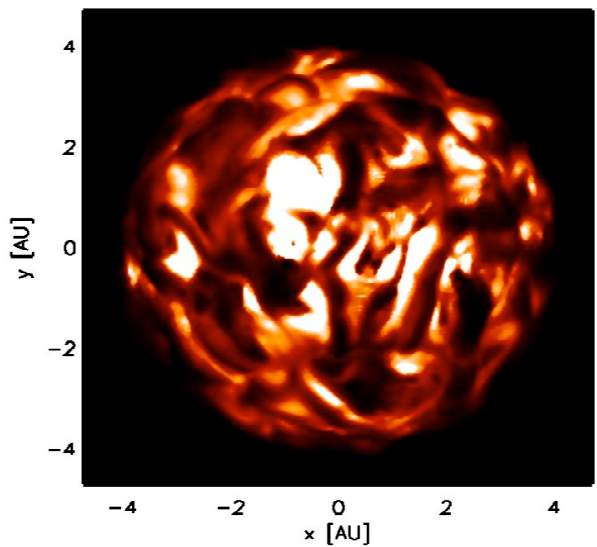}\hspace{2pc}%
\begin{minipage}[b]{16pc}\caption{\label{map}Synthetic map of intensity (the range is [0 -- 230000]
        erg/s/cm$^2$/\AA ) in the Gaia $G$ band \cite{2010A&A...523A..48J}. Figure taken from \cite{2011A&A...528A.120C}.}
\end{minipage}
\end{figure}

\begin{figure}[h]
  \begin{tabular}{c}
   \includegraphics[width=0.49\hsize]{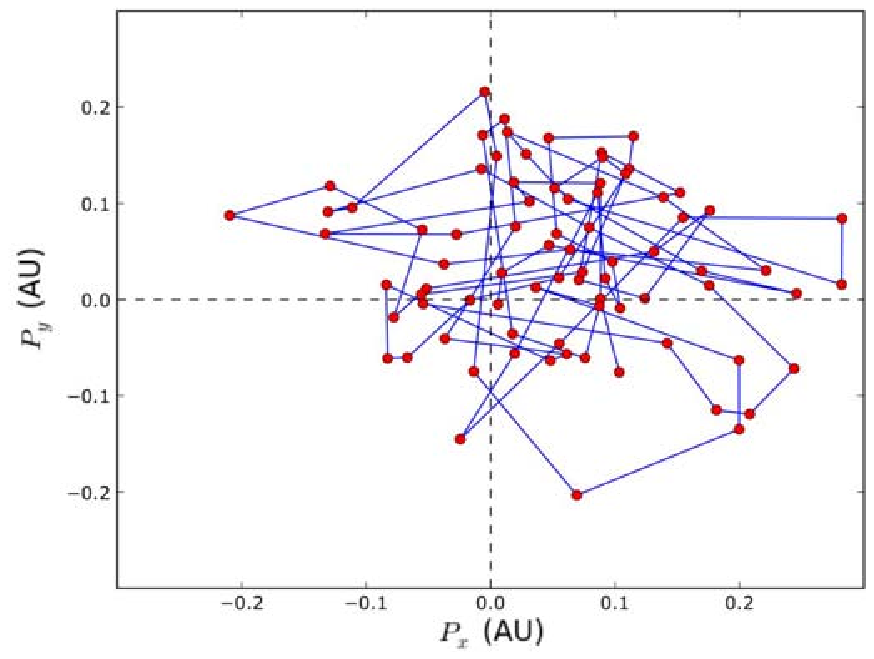}
   \includegraphics[width=0.48\hsize]{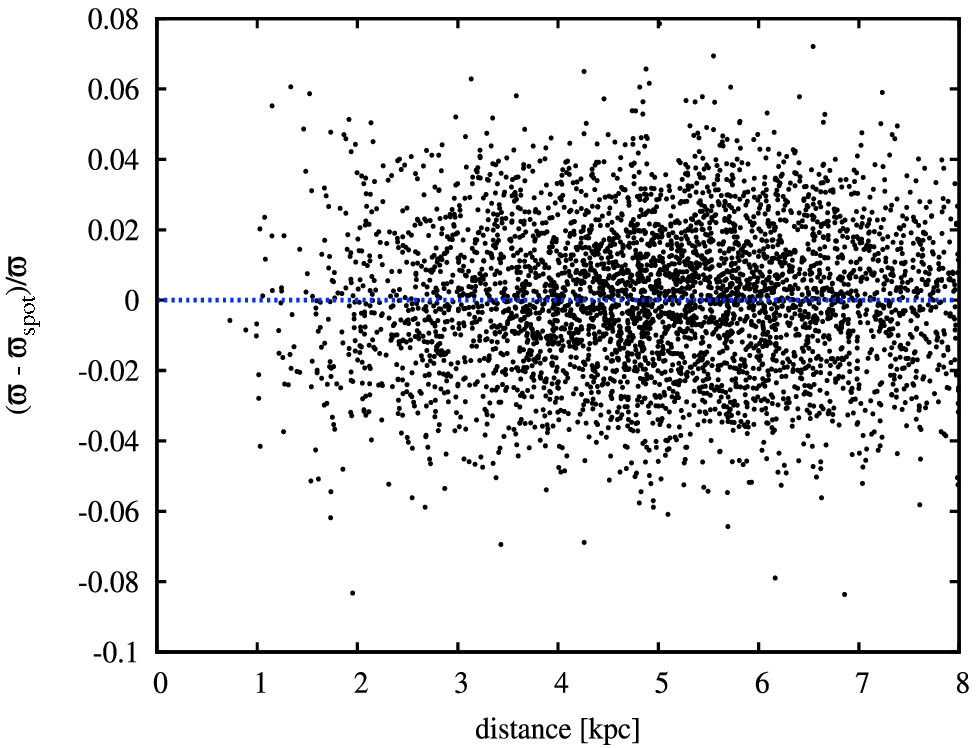}
\end{tabular}
     \caption{\label{photo}\emph{Left:} photocenter displacement as a function of time for a RSG simulation. Each point is a snapshot
23 days apart for a total of 5 years of simulation (comparable to the duration of the Gaia mission). \emph{Right:} relative
difference between the parallaxes computed with and without the photocentric motion of the left panel, as a function
of the distance. Figures taken from \cite{2011A&A...528A.120C}.}
\end{figure}

The simulated surface of RSGs (Fig.~\ref{map}) displays high-contrast structures with spots up to 50 times brighter than the dark ones with strong changes over some weeks. This aspect is connected with the underlying granulation pattern, but also with dynamical effects such as shocks and waves which dominate at optical depths smaller than~1 \cite{2010A&A...515A..12C}. The convective-related surface structures in the Gaia $G$ band affects strongly the position of the photocenter and cause temporal fluctuations (Fig.~\ref{photo}, left). The photocenter excursion is
large, since it goes from 0.005 to 0.3~AU over 5 years of simulation and it is on average 0.132~AU (i.e., $\sim3\%$ of the stellar radius). This value corresponds to a photocenter excursion of $\sim$10 $\mu$as for a star at 1 kpc, which is comparable to the Gaia accuracy of $\sim$10$\mu$as for stars brighter than 10.\\
Chiavassa et al. \cite{2011A&A...528A.120C} calculated the Gaia parallax using Gaia Object Generator v7.0, GOG\footnote{http://gaia-gog.cnes.fr} \cite{2010hsa5.conf..415I} for RSGs with surface brightness asymmetries from RHD simulations ($\varpi_{\rm{spot}}$) and without ($\varpi$). Figure~\ref{photo} (right) shows that there is a systematic error of a few percent. This systematic error may be up to 15 times the formal error $\sigma_{\varpi}$ \cite{2011A&A...528A.120C}.
Although this error should be used to revise $\sigma_{\varpi}$, there is little hope to be able to correct the Gaia parallaxes of RSGs from this parallax error, without knowing the run of the photocentric shift for each considered star. Thus, it might be of interest to monitor the photocentric deviations of a number of selected RSGs during the Gaia mission using interferometry and/or spectroscopy to obtain valuable information on the photocentric position and correct the resulting parallaxes using RHD predictions.

\section{Conclusions and perspectives}

3D hydrodynamical simulations of surface convection are a very important tool to account for convective shifts correction on the kinematic radial velocities, parallaxes and photometry. \\

With the increasing computer power, it is now possible to compute 3D model grids rather easily on reasonable timescale. In this framework, the 3D local model grid done with {\sc Stagger-Code} (see contribution of Remo Collet in this Volume) will be available very soon together with the synthetic spectra in RVS domain (before Gaia launch date, mid 2013). The grid will include about 130 models with $T_{\rm{eff}}$ ranging from 4000 to 6500 K, $\log g$ from 1.5 to 5 and [Fe/H]=0.0, $-$1.0, $-$2.0, $-$3.0. These models will not be enough to represent all the stars observed by Gaia and thus, following \cite{2011EAS....45..195J}, we plan to apply 3D convective corrections to spectroscopic radial velocities obtained with less sophisticated models and templates. 

\ack
The authors thank the Rechenzentrum Garching (RZG) for providing the computational resources necessary for this
work. 

\section*{References}
\bibliographystyle{iopart-num}
\bibliography{iopart-num}

\end{document}